\documentstyle[12pt,epsf]{article}
\input epsf.sty
\begin{document}
\title{Hadronic Matter with Internal Symmetries and its Consequences:
An Expanding Hadronic Gas
\thanks{Invited lecture at the "Advanced NATO Workshop: Hot Hadronic
Matter: Theory and Experiment", Divonne-les-Bains, June 27 -
July 1, 1994.}}
\author{Ludwik Turko
\\ Institute of Theoretical Physics, University of Wroc{\l}aw,
\\ Pl.Maksa Borna 9, 50-204  Wroc{\l}aw, Poland}
\date{}
\maketitle
\newcommand{\RRR}{R\hspace{-0.4cm}R}
\newcommand{\NNN}{N\hspace{-0.31cm}N}
\newcommand{\ZZZ}{Z\hspace{-0.4cm}Z}
\newcommand{\ds}{\displaystyle}
\newcommand{\un}{\underline}
\newcommand{\be}{\begin{equation}}
\newcommand{\ee}{\end{equation}}
\newcommand{\ba}{\begin{array}}
\newcommand{\ea}{\end{array}}
\newcommand{\bd}{\begin{description}}
\newcommand{\ed}{\end{description}}
\newcommand{\po}{Poincar\'{e}}
\def\NN{{\cal N}}
\begin{abstract}
We consider an ideal gas of massive hadrons in thermal and chemical
equilibrium. The gas expands longitudinally in accordance with Bjorken
law. Strangeness and baryon number conservation is taken into account.
This gas has different features as compared to the pion gas.
\end{abstract}

\section*{ }
Let us consider a heavy ion collision which goes through the intermediate
state of an hadronic matter being in the thermal and chemical equilibrium.
This matter can be considered as a gas which expands longitudinally
according to a hydrodynamical evolution. In the simplest case the
hadronic matter is supposed to be made of non - interacting massless
pions [1]. However an assumption of a thermodynamical equilibrium leads to
the conclusions that a system under consideration should consist of all
species of hadrons allowed by conditions of the equilibrium.  We are
going to consider an ideal gas consisting of all species of hadrons up to
$\Omega^{-}$ baryon. These realistic hadrons need that conservation
laws due to an internal symmetries should be taken into account.

The aim of this talk is to present some of pecularities of this more
realistic hadronic gas as compared to the simple pion gas. Some of
these comparison were published before to study $J/\Psi$ suppression
patterns [2].

For an ideal hadron gas in thermal and chemical equilibrium, which consists of
$l$ species of particles, energy density $\epsilon$, baryon number density
$n_{B}$, strangeness density $n_{S}$ and entropy density $s$ read
($\hbar=c=1$ always)
$$
\epsilon = { 1 \over {2\pi^{2}}} \sum_{i=1}^{l} (2s_{i}+1) \int_{0}^{\infty}
{ { dpp^{2}E_{i} }
\over { \exp \left\{ {{ E_{i} - \mu_{i} } \over T} \right\}
+ g_{i} } } \ ,
\eqno(1a)
$$
$$
n_{B}={ 1 \over {2\pi^{2}}} \sum_{i=1}^{l} (2s_{i}+1) \int_{0}^{\infty}
{ { dpp^{2}B_{i} }
\over { \exp \left\{ {{ E_{i} - \mu_{i} } \over T} \right\}
+ g_{i} } } \ ,
\eqno(1b)
$$
$$
n_{S}={1 \over {2\pi^{2}}} \sum_{i=1}^{l} (2s_{i}+1) \int_{0}^{\infty}
{ { dpp^{2}S_{i} }
\over { \exp \left\{ {{ E_{i} - \mu_{i} } \over T} \right\}
+ g_{i} } } \ ,
\eqno(1c)
$$
$$
s={1 \over {6\pi^{2}T^{2}} } \sum_{i=1}^{l} (2s_{i}+1) \int_{0}^{\infty}
{ {dpp^{4}} \over { E_{i} } } { { (E_{i}
- \mu_{i}) \exp \left\{ {{ E_{i} - \mu_{i} } \over T}
\right\} } \over { \left( \exp \left\{ {{ E_{i} - \mu_{i} }
\over T} \right\} + g_{i} \right)^{2} } }\ ,
\eqno(1d)
$$
where $E_{i}= ( m_{i}^{2} + p^{2} )^{1/2}$ and
$m_{i}$, $B_{i}$, $S_{i}$, $\mu_{i}$, $s_{i}$ and $g_{i}$ are the mass,
baryon number, strangeness, chemical potential, spin and
a statistical factor of specie $i$ respectively (we treat
an antiparticle as a different specie). And $\mu_{i} =
B_{i}\mu_{B} +
S_{i}\mu_{S}$, where $\mu_{B}$ and $\mu_{S}$ are overall baryon number and
strangeness chemical potentials respectively.

In general, all densities of the left sides of eqs.1a-d are functions of time
during the cooling of the hadron gas. This means that gas parameters such as
temperature and chemical potentials should be functions of time. We would like
to obtain these functions explicite.
This can be done by solving numerically the system
of equations for entropy density $s$, baryon number density $n_{B}$ and
strangeness density $n_{S}$, with $s$, $n_{B}$ and $n_{S}$ given as
time dependent quantities.
{}From the Bjorken model we have the following solution for
the longitudinal expansion [1,5]
$$
s(t)= { {s_{0}t_{0}} \over t } \;,\;\;\;\;\;\;\;\;\;
n_{B}(t)= { {n_{B}^{0}t_{0}} \over t } \ ,
\eqno(2)
$$
where $s_{0}$ and $n_{B}^{0}$ are initial densities of the entropy and the
baryon number respectively.
The overall strangeness is equal to zero during all the evolution.
So to solve (1b,c,d) with $s$ and $n_{B}$ given by (2) and $n_{S}=0$,
we need to know initial values $s_{0}$ and $n_{B}^{0}$.
To estimate initial baryon number density $n_{B}^{0}$ we use experimental
results of [3].

These results are for S-S collisions, but since there are no data on
baryon multiplicities for heavier nuclei we have to evaluate them in
some way.  We assume that the baryon multiplicity per unit rapidity in
the CRR is proportional to the number of participating nucleons. For a
sulphur-sulphur collision we have $dN_{B}/dy \cong 6$ and 64
participating nucleons [3].  For an O-U collision we can roughly estimate
the number of participating nucleons at $16+58=74$. The second factor
of the sum has been obtained by the following assumption: since an
oxygen nucleon is much smaller than an uranium one, we can approximate
the part of the uranium, through which the oxygen passes, by the
cylinder of the volume equal to $\pi R_{O}^{2} \cdot 2R_{U}$. The same
procedure can be applied to the S-U case. Here we obtain $32+93=125$
participants. Therefore, we have $dN_{B}/dy \cong 7$ and $dN_{B}/dy
\cong 11.7$ for O-U and S-U collisions respectively. Having
taken the initial volume in the CRR equal to $\pi R_{A}^{2} \cdot
1$ fm, we arrive at $n_{B}^{0} \cong 0.25 \; {\rm fm^{-3}}$ for both cases.

To find $s_{0}$, first we have to solve (1a,b,c) with respect to $T$,
$\mu_{S}$ and $\mu_{B}$, where we put $\epsilon = \epsilon_{0}$ and so
on.  For $\epsilon_{0}$ we have taken estimates given in [4].

As a result, we have obtained $T_{0} \cong 212$ MeV, $\mu_{S}^{0} \cong
34.6$ MeV and $\mu_{B}^{0} \cong 133$ MeV for the O-U collision
($\epsilon_{0}= 2.5 \; {\rm GeV/fm^{3}}$) and $T_{0} \cong 209.1$ MeV,
$\mu_{S}^{0} \cong 36.7$ MeV and $\mu_{B}^{0} \cong 143.3$ MeV for the
S-U collision ($\epsilon_{0}= 2.3 \; {\rm GeV/fm^{3}}$). Then, from
(1d) we have $s_{0} \cong 13.68 \; {\rm fm^{-3}}$ for O-U and $s_{0}
\cong 12.74 \;
{\rm fm^{-3}}$ for S-U.  Now, having put (2) and $n_{S}= 0$ into
(1b,c,d), we can solve them numerically to obtain $T$, $\mu_{S}$ and
$\mu_{B}$ as functions of time.  Our results are presented in fig.1,
where solid, long-dashed and dashed lines mean the temperature, the
strangeness chemical potential and the baryon chemical potential
respectively. Fig.1 shows results for $S - S$ collision initial
conditions. The time scale is chosen in a way which enables the
temperature to reach the freez-out at 140 MeV.  This corresponds to the
freez-out time equal to $t_{f.o.} \cong 10.4$ fm .

 The solution for the temperature
function can be approximated by an expression of the form
$$
T(t) = T_{0} \cdot \left( {1 \over t} \right)^{a}\ ,
\eqno(3)
$$

where

$T_0 = 212.4 MeV ,  a = 0.178 $ for $O - U$ system

and

$T_0 = 209.7 MeV ,  a = 0.179 $ for $S - S$ system.

We can see that all above expressions have the form known from the solution
for the longitudinal expansion of a baryonless gas with the sound velocity
constant, namely [5]
$$
T(t) = T_{0} \cdot \left( {1 \over t} \right)^{ c_{s}^{2} }\ ,
\eqno(4)
$$
where $c_{s}$ is the sound velocity and we put the initial time $t_{0}$
equal to 1 fm.
We have checked that for $n_{B} = 0$ results for the temperature function
are very similar to those in eq.3.
And the following approximations of the temperature function hold
$$
T(t) = 215.2 \cdot \left( {1 \over t} \right)^{0.180}
\eqno(4a)
$$
$$
T(t) = 208.9 \cdot \left( {1 \over t} \right)^{0.172}.
\eqno(4b)
$$

We can see that the power in eqs. 3 and 4 is around 0.18. Therefore a
question arises: is this value connected in any way with the sound velocity
of the hadron gas? For the baryonless case the answer is positive. We have
checked this computing straightforward $c_{s}^{2}=s/(T { {\partial s} \over
{\partial T} })$, which is the value of the sound velocity squared for
a baryonless gas [5]. The results are presented in fig.2.
For comparision, the square of the sound velocity of a pion
gas is also depicted (dashed lines). We can see that in the range of the
temperature 200-140 MeV
the square of the speed of sound equals 0.17-0.18 indeed. We have also found
a region of the temperature where the sound velocity decreases as the
temperature increases. Nevertheless, the condition for the stability of
the expansion [6], $d/dT(sc_{s}/T) > 0$ is still valid because $sc_{s}/T$
is an increasing function of time.

\section*{Acknowledgments}
It is pleasure to thank Prof. J.Rafelski for his kind invitation to
contribute to this workshop and for his warm hospitality.

\newpage

\section*{Figure Captions}

\bd

\item{Fig.1.} Dependence of chemical potentials and temperature on
time for $S - S$ collision initial condition. Solid,
long-dashed and dashed lines mean the temperature, the
strangeness chemical potential and the baryon chemical potential
respectively.

\item{Fig.2.} Time - dependence of the square of the sound velocity for the
expanding hadron gas. Long-dashed and solid lines mean the square of
the sound velocity for the pion gas and for the baryonless hadron case
respectively.

\ed

\newpage
{\epsfbox{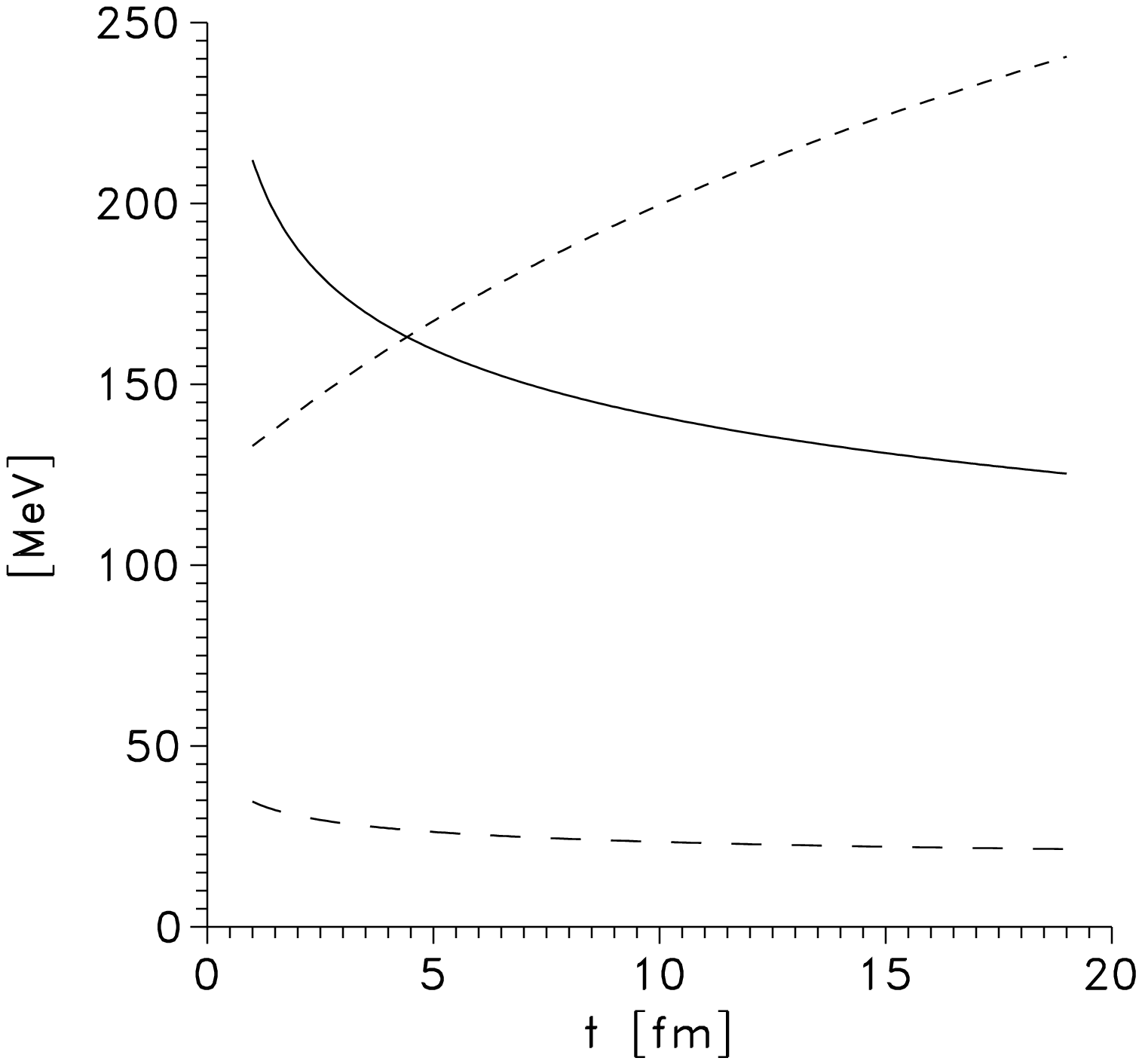}}

\newpage

{\epsfbox{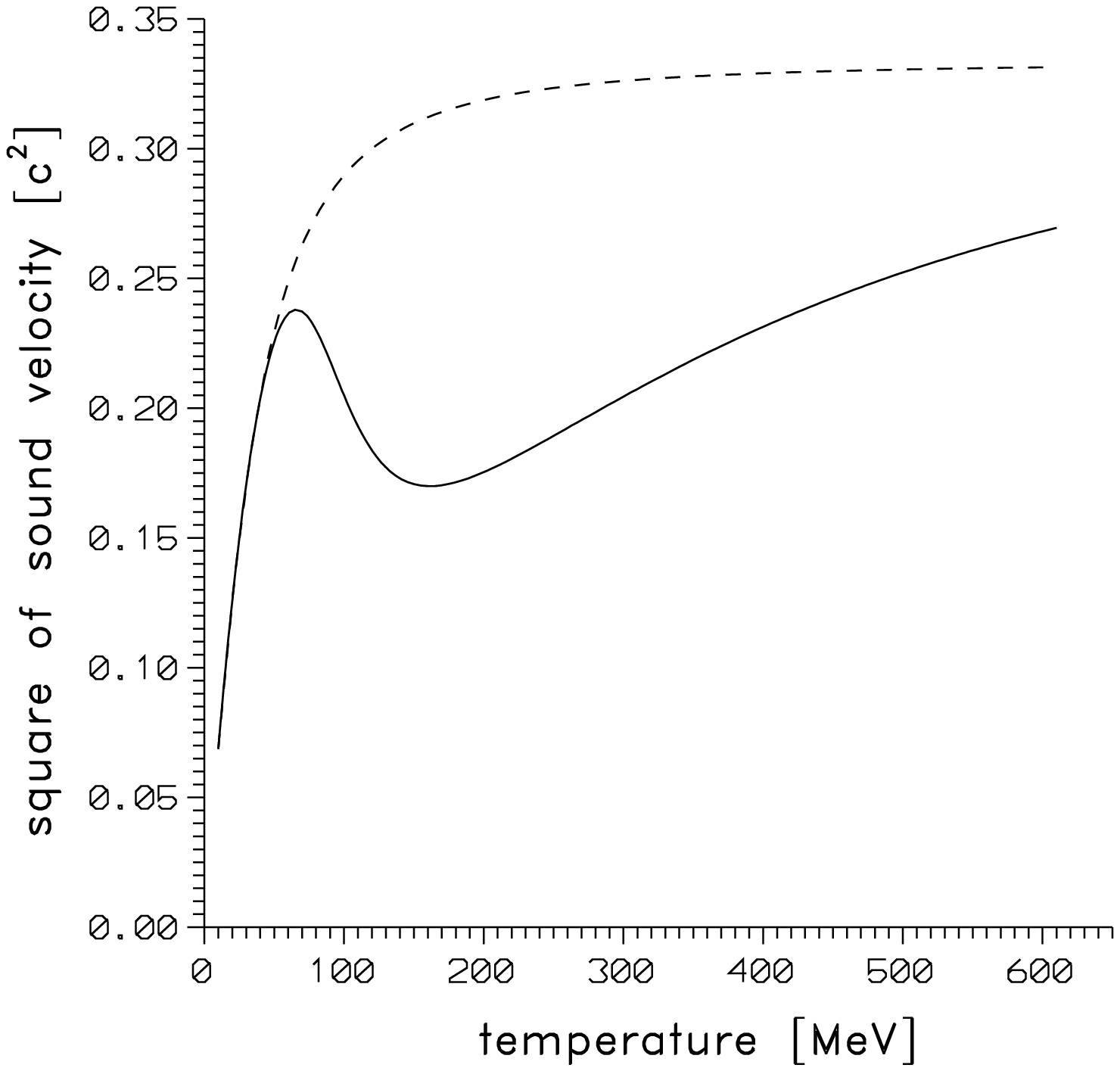}}

\end{document}